# A vision-intelligent framework for mapping the genealogy of vernacular architecture

Xuan Xue[1, †], Yaotian Yang[2, †], Zihui Tian[1, †], T.C. Chang[3], Chye Kiang Heng[2, *]

[1] Department of Architecture, National University of Singapore, Singapore, Singapore
[2] Department of Chemical Engineering, Tsinghua University, Beijing, China
[3] Department of Geography, National University of Singapore, Singapore, Singapore
[†] These authors contributed equally: Xuan Xue, Yaotian Yang, Zihui Tian
[*] e-mail: akihck@nus.edu.sg

## Abstract

The study of vernacular architecture involves recording, ordering, and analysing buildings to probe their physical, social, and cultural explanations. Traditionally, this process is conducted manually and intuitively by researchers. Because human perception is selective and often partial, the resulting interpretations of architecture are invariably broad and loose, often lingering on form descriptions that adhere to a preset linear historical progression or crude regional demarcations. This study proposes a research framework by which intelligent technologies can be systematically assembled to augment researchers' intuition in mapping or uncovering the genealogy of vernacular architecture and its connotative socio-cultural system. The vision-intelligent framework is predicated on the premise that the interpretation of historical artefacts is a process of identifying a function f from a discrete data set. This process begins with the formulation of a hypothesis, followed by three phases - *collection*, *categorisation*, and *interpretation* - and culminates in the *verification* of the hypothesis and the *articulation* of final explanations. We employ this framework to examine the stylistic classification of 1,277 historical shophouses in Singapore's Chinatown. Findings extend beyond the chronological classification established by the Urban Redevelopment Authority of Singapore in the 1980s and 1990s, presenting instead a phylogenetic network to capture the formal evolution of shophouses across time and space. The network organises the shophouse types into ***nine*** distinct clusters, revealing concurrent evidences of cultural evolution and diffusion. Moreover, it provides a critical perspective on the multi-ethnic character of Singapore shophouses by suggesting that the distinct cultural influences of different ethnic groups led to a pattern of parallel evolution rather than direct convergence. Our work advances a quantitative genealogy of vernacular architecture, which not only assists in formal description but also reveals the underlying forces of development and change. It also exemplified the potential of collaboration between studies in vernacular architecture and computer science, demonstrating how leveraging the strengths of both fields can yield remarkable insights.

## 1. Introduction

Vernacular architecture is one of the most ubiquitous forms of material culture that can facilitate an understanding of people and their history and society[1]. Vernacular architecture is also one of the most underrated cultural heritages, often overshadowed by monumental or iconic structures and constantly threatened by destruction under relentless waves of rapid urbanisation and modernization[2]. The study of vernacular architecture has been a common ground for antiquarians, historians, geographers, folklorists, archaeologists, anthropologists, architects and more[3]. For years, they have made great strides to record, order and analyse buildings, so as to better understand their environmental, technological, social, cultural and aesthetic characteristics and to inform future conservation and development practices[4].

Understanding past spatio-temporal patterns and processes of building change ranks among the key goals of vernacular architecture studies. Recognising such processes relies heavily on building classification. Traditionally, this work is conducted manually and intuitively by researchers, entailing extensive labour which might often overlook intricate architectural types or variants. Such a situation might be tolerable from an art historical consideration because its interpretation of vernacular architecture is intrinsically selective, depending on the compatibility of form with contemporary aesthetics[5]. However, from a social scientific perspective, "the attempt to account for all buildings is fundamental"[6], and the conditions noted above may impede the credibility of building genealogy, undermining efforts in understanding societal and cultural change[7].

In recent years, the rapid development of spatial data acquisition, computer vision, and scientific data analysis has showcased great potential to move the research paradigm of vernacular architecture towards efficiency, objectivity and profundity. The dividends are manifold. Firstly, the rise of varied technologies of spatial data acquisition, such as geo-information systems, 3D scanning, digital archives, and more, have produced vast amounts of data relating to vernacular architecture. Services like Google Street View have provided online access to images that cover most parts of the world and anything photographable from a public street. This democratic rather than discriminative record of extant buildings ensures high credibility [8] and are also publicly accessible, underscoring the great potential to liberate researchers from intense data collection fieldwork[9].

---

[1] Henry Glassie, *Folk Housing in Middle Virginia: A Structural Analysis of Historic Artifacts*, First Edition, First (Knoxville: Univ Tennessee Press, 1976); Paul Oliver, "Vernacular Know-How," *Material Culture* 18, no. 3 (1986): 113–26; Victor Buchli, *An Anthropology of Architecture* (London; New Delhi; New York; Sydney: Bloomsbury Publishing Plc, 2013).

[2] Paul Oliver, *Built to Meet Needs: Cultural Issues in Vernacular Architecture* (Oxford: Architectural Press, 2006).

[3] Roderick J. Lawrence, "The Interpretation of Vernacular Architecture," *Vernacular Architecture* 14, no. 1 (1983): 19–28, https://doi.org/10.1179/vea.1983.14.1.19; Paul Oliver, "Why Study Vernacular Architecture? (1978)," in *Built to Meet Need: Cultural Issues in Vernacular Architecture* (1978; repr., Oxford: Architectural Press, 2006), 3–16.

[4] Lawrence, "The Interpretation of Vernacular Architecture."

[5] Henry Glassie, "Eighteenth-Century Cultural Process in Delaware Valley Folk Building," *Winterthur Portfolio* 7 (January 1972): 29–57, https://doi.org/10.1086/495804; Dell Upton, "The Power of Things: Recent Studies in American Vernacular Architecture," *American Quarterly* 35, no. 3 (1983): 262–79, https://doi.org/10.2307/2712651; Lawrence, "The Interpretation of Vernacular Architecture."

[6] Glassie, "Eighteenth-Century Cultural Process in Delaware Valley Folk Building," 31.

[7] Glassie, *Folk Housing in Middle Virginia*.

Secondly, technological advancements in computer vision have developed varied image-processing algorithms that can be used to solve the problems of architectural classification[10]. Early works focused on solving the repetitive labour in architectural classification tasks rather than formulating new architectural typological systems. They used simpler machine-learning algorithms to classify images with manually extracted features (such as edges, textures, corners, etc.). For example, Shalunts et al. (2011)[11] proposed classifying a small group of images of building façade windows with three pre-determined architectural styles: Romanesque, Gothic and Baroque. By using the algorithm Scale Invariant Feature Transform to extract local features, the study completed the task with a high accuracy of 95.16%. The rise of deep learning in the 2010s, specifically convolutional neural networks (CNNs), has allowed for more sophisticated and automated image classification tasks. Works that probed the architectural typological systems started to emerge. Based on the deep convolutional neural network (DCNN), Sun et al. (2022)[12] re-examined the stylistic system of residential buildings in Cambridge, U.K., formulated by Lindenthal and Johnson (2021)[13]. The study demonstrates the correlation between style and age, which manifests most prominently in the architectural feature of the window. This discussion on the AI-augmented formal taxonomy is preliminary but already shows great potential for deeper investigation.

Finally, scientific data analysis packages have developed many mathematical models to describe and interpret the relationship among taxa[14]. Their application to archaeology and linguistics is widely seen but rarely applied to vernacular architecture. One problem is that the interpretation of vernacular architecture that underpins much of the study is seldom elaborated explicitly[15]. In contrast, other domains within cultural studies have evolved to establish mathematical models based on certain propositions about culture[16]. For example, archaeologists employ phylogenetic approaches to model change in artefact forms across time and space, like stone tools[17], pottery[18], carpets[19] and more. This is based on the proposition that "cultural change is systematic rather than capricious and … that an important basis for the systematic behaviour of culture is its continuous transmission through the agency of person-to-person contact."[20] Transferring these propositions and numerical methods to vernacular architecture studies is promising. On the one hand, it contributes to validating existing theories about vernacular architecture; on the other hand, it can yield results that are theoretically justifiable, reproducible, and quantitatively defensible.

---

[8] Filip Biljecki and Koichi Ito, "Street View Imagery in Urban Analytics and GIS: A Review," *Landscape and Urban Planning* 215 (November 1, 2021): 104217, https://doi.org/10.1016/j.landurbplan.2021.104217; Mike Christenson, "Viewpoint: 'From the Unknown to the Known': Transitions in the Architectural Vernacular," *Buildings & Landscapes* 18, no. 1 (2011): 1–13, https://doi.org/10.1353/bdl.2011.0005.

[9] Hannah M. Badland et al., "Can Virtual Streetscape Audits Reliably Replace Physical Streetscape Audits?," *Journal of Urban Health* 87, no. 6 (December 1, 2010): 1007–16, https://doi.org/10.1007/s11524-010-9505-x; Adam Berland and Daniel A. Lange, "Google Street View Shows Promise for Virtual Street Tree Surveys," *Urban Forestry & Urban Greening* 21 (January 1, 2017): 11–15, https://doi.org/10.1016/j.ufug.2016.11.006.

[10] Małgorzata B. Starzyńska et al., "Computer Vision-Based Analysis of Buildings and Built Environments: A Systematic Review of Current Approaches" (arXiv, August 1, 2022), https://doi.org/10.48550/arXiv.2208.00881.

[11] Gayane Shalunts, Yll Haxhimusa, and Robert Sablatnig, "Architectural Style Classification of Building Facade Windows," in *Advances in Visual Computing*, ed. George Bebis et al. (Berlin, Heidelberg: Springer, 2011), 280–89, https://doi.org/10.1007/978-3-642-24031-7_28.

[12] Maoran Sun et al., "Understanding Architecture Age and Style through Deep Learning," *Cities* 128 (September 2022): 103787, https://doi.org/10.1016/j.cities.2022.103787.

[13] Thies Lindenthal and Erik B. Johnson, "Machine Learning, Architectural Styles and Property Values," *The Journal of Real Estate Finance and Economics*, July 13, 2021, https://doi.org/10.1007/s11146-021-09845-1.

[14] Daniel H. Huson and David Bryant, "The SplitsTree App: Interactive Analysis and Visualization Using

Drawing on extant scholarly endeavours, this study aims to advance computational vernacular studies by guiding human intuition with vision-intelligent technologies in formulating the genealogy of vernacular architecture. By incorporating *collection*, *categorisation* and *interpretation*, our proposed framework provides researchers with a systematic digital toolkit to examine built forms in dispersed geographical contexts, in order to derive justifiable maps of relatedness.

The Singapore shophouse, a quintessential urban vernacular architecture[21], has been selected as the subject of our scholarly investigation. The study created and analysed a georeferenced image set of surviving 19th and 20th-century shophouses in downtown Singapore. The quantitative modelling results indicate that the formal

---

Phylogenetic Trees and Networks," *Nature Methods* 21, no. 10 (October 2024): 1773–74, https://doi.org/10.1038/s41592-024-02406-3; Øyvind Hammer, David A.T. Harper, and Paul D. Ryan, "Past: Paleontological Statistics Software Package for Educaton and Data Anlysis," *Palaeontologia Electronica* 4, no. 1 (2001): 1–9.

[15] Lawrence, "The Interpretation of Vernacular Architecture"; Dell Upton, "The VAF at 25: What Now?," *Perspectives in Vernacular Architecture* 13, no. 2 (2007 2006): 7–13, https://www.jstor.org/stable/20355380.

[16] Carl P. Lipo et al., eds., *Mapping Our Ancestors: Phylogenetic Approaches in Anthropology and Prehistory* (New York: Routledge, 2006), https://doi.org/10.4324/9780203786376.

[17] K. A. Robson Brown, "Systematics and Integrated Methods in the Modelling of the Pre-Modern Human Mind," in *Modelling the Early Human Mind*, ed. Kathleen R. Gibson and Paul Mellars (McDonald Institute for Archaeological Research, University of Cambridge, 1996), 107–22.

[18] Mark Collard and Stephen Shennan, "Ethnogenesis versus Phylogenesis in Prehistoric Culture Change: A Case-Study Using European Neolithic Pottery and Biological Phylogenetic Techniques.," in *Archaeogenetics: DNA and the Population Prehistory of Europe*, ed. Lord Colin Renfrew and Katie Boyle (McDonald Institute for Archaeological Research, 2000), 89–97, https://www.semanticscholar.org/paper/Processes-of-culture-change-in-prehistory%3A-a-case-Shennan-Collard/815f09a171445384d7571ccc0abe460946312546.

[19] Jamshid Tehrani and Mark Collard, "Investigating Cultural Evolution through Biological Phylogenetic Analyses of Turkmen Textiles," *Journal of Anthropological Archaeology* 21, no. 4 (December 2, 2002): 443–63, https://doi.org/10.1016/S0278-4165(02)00002-8.

[20] Albert C. Spaulding, *Prehistoric Cultural Development in the Eastern United States*, The Bobbs-Merrill Reprint Series in the Social Sciences (Indianapolis: Bobbs-Merrill, 1955), 14.

[21] Ho Yin Lee, "The Singapore Shophouse: An Anglo-Chinese Urban Vernacular," in *Asia's Old Dwelling: Tradition, Resilience, and Change*, ed. G. Knapp Ronald (New York: Oxford University Press, 2003), 115–34; T.C. Chang and Peggy Teo, "The Shophouse Hotel: Vernacular Heritage in a Creative City," *Urban Studies* 46, no. 2 (February 2009): 341–67, https://doi.org/10.1177/0042098008099358; Julian Davison, *Singapore Shophouse* (Singapore: Talisman Publishing, 2010).

change of shophouses across time and space is best captured by a phylogenetic network, which organises the shophouse types into **nine distinct clusters.** In turn, we also show how this classification reveals concurrent signals of cultural evolution and diffusion. Here, cultural evolution manifests as a transition in architectural styles from the traditional to the modern, while cultural diffusion—particularly during periods of economic prosperity—is characterised by the emergence of diverse ethnic styles, indicating cultural exchange among different ethnicities, alongside competition and integration.

The report begins with some conceptual background and literature review on how we might 'compute the vernacular'. The goal is to develop a computational framework to facilitate visual intelligence in interpreting historical building artefacts. This is followed by some contextual information on Singapore and its shophouse vernacular. The data is then presented in a three-fold discussion highlighting the processes of *collection*, *categorisation* and *interpretation*. The study ends by verifying the hypothesis and thinking through the wider applications of our work to other architectural contexts, practices and empirical settings.

## 2. Computing the Vernacular

While it is tempting to imagine vernacular architecture as changeless, we must acknowledge that change is inherent in the nature of the vernacular[22]. Classification is also indispensable in describing the change in vernacular buildings and landscapes across time and space.

Traditional approaches to classification rely on the researchers' intuition established through constant re-experiencing of identical or similar artefacts. Fred Knifen's (1965) pioneering work on settlement geography in the U.S. involves the recording and classification of folk housing. He lamented, "There are times in the field when a deluge of variant or new types and combinations leads to a state of utter confusion on the part of the observer. There is then no alternative to stopping for as long as is necessary to identify the pieces and fit them into place"[23]. Because human perception is selective and partial, the resulting interpretation of architecture is necessarily broad and loose, often lingering on form descriptions that follow a linear historical progression or a crude regionalist model.

The *Encyclopaedia of Vernacular Architecture of the World*, edited by Oliver Paul (1997), is oft-regarded the most comprehensive reference work for international vernacular architecture studies till today, bringing together the vernacular traditions of over 2,000 different cultures from around the world[24]. Its categorisation of vernacular buildings is based on cultural regions delineated by existing global mappings. While reflecting the nature of vernacular architecture generally, this system is unable to capture the cross-cultural or geographical particularities of vernacular traditions[25].

[22] Lawrence, "The Interpretation of Vernacular Architecture," 26; Dell Upton and John Michael Vlach, "Introduction," in *Common Places: Readings in American Vernacular Architecture* (Athens and London: The University of Georgia Press, 1986), xx; Lindsay Asquith, "Lessons from the Vernacular: Integrated Approaches and New Methods for Housing Research," in *Vernacular Architecture in the 21st Century* (Taylor & Francis, 2005), 129.

[23] Fred Kniffen, "Folk Housing: Key to Diffusion," *Annals of the Association of American Geographers* 55, no. 4 (December 1, 1965): 556, https://doi.org/10.1111/j.1467-8306.1965.tb00535.x.

Researchers have acknowledged the gravity of the situation, and efforts to deepen the study are underway. Henry Glassie is one such vanguard scholar who advocated precise definitions and systematic coding of artefacts under investigation. Despite no computational technique involved, his work *Folk Housing in Middle Virginia: A Structural Analysis of Historic Artifacts* (1976) adopts a rigorous approach to formal taxonomy to investigate 338 real houses. Each building is closely examined based on a pre-defined artefactual grammar, which describes the recurring building elements and their syntactic arrangement to generate a genealogy of building types. Informed by structuralism, his study aims to move beyond the building forms and uncover the human ideas behind them. He explained the architectural change in Middle Virginia through the makers' design competence, "he is reliant not on one original, but on a competence constructed out of numerous originals. He labours within tight lines of correctness, but his design competence allows for a variable range of actions."[26]

Drawing on Glassie's work, we propose a computational framework in which researchers can use visual intelligence tools to guide their intuition concerning vernacular buildings, helping them categorise buildings and also unpack deeper structures that govern formal changes. Sitting at the core of this framework is a belief that **the interpretation of historical artefacts is a process of finding a function *f* from a discrete data set.** This epistemological stance, visualised in **Figure 1**, enables a computational study at a scale and depth that traditional methods cannot achieve. We propose that the research framework starts with establishing a cultural hypothesis on architectural change, followed by the three phases of data ***collection***, ***categorisation***, and ***interpretation***, concluding with verifying the hypothesis and providing final explanations.

In the ***collection*** phase, researchers must prepare the building image set and formulate the feature rules based on the established hypothesis. As mentioned earlier, the image set generated from various data acquisition technologies, can perform with high integrity and accuracy. Each image in the image set can be read as a discrete point within space, and these points will converge into clusters during the subsequent categorisation phase. By labelling a certain amount of building images with feature rules, we can train the supervised model to auto-detect all formal features of each building image. The key contribution of computer vision and deep learning is its ability to automate the observation and recording of features for each building, representing these features using ‘strings’. Building images with similar strings will be automatically grouped into the same cluster.

[24] Paul Oliver, *Encyclopedia of Vernacular Architecture of the World*, 3 vols. (Cambridge: Cambridge University Press, 1997).
[25] Marcel Vellinga, Paul Oliver, and Alexander Bridge, *Atlas of Vernacular Architecture of the World* (Routledge, 2007), XV.
[26] Glassie, *Folk Housing in Middle Virginia*, 67.

The ***categorisation*** phase concludes with a set of equivalent building clusters represented with X(y). These equivalent clusters will ultimately be connected by structural lines informed by function *f* in the ***interpretation*** phase. The function *f* is derived from the analytical models with certain cultural hypotheses. Indebted to the collaborative endeavour between mathematicians and social scientists, these models can be found in their developed scientific data analysis packages. By taking the building clusters represented with X(y) as model input, we get a candidate description *f*′(X(y)). In many cases, the candidate explanation is not perfectly applicable when referring to historical archives of the studied object. This prompts the researcher to revise the hypothesis and alter the model. The iterative process may need to be repeated multiple times before arriving at a satisfactory explanation f(X(y)).

We suggest integrating these well-established techniques from remote sensing, deep learning and statistics, which offer a rigorous and productive method to support the researcher’s work. This vision-intelligent research framework, we argue, establishes a rationalist epistemological stance on vernacular architecture studies. The next section provides a brief background on Singapore’s shophouse vernacular before we demonstrate in the subsequent section how the framework has been successfully used to document and explain formal changes in Singapore shophouses across time and space.

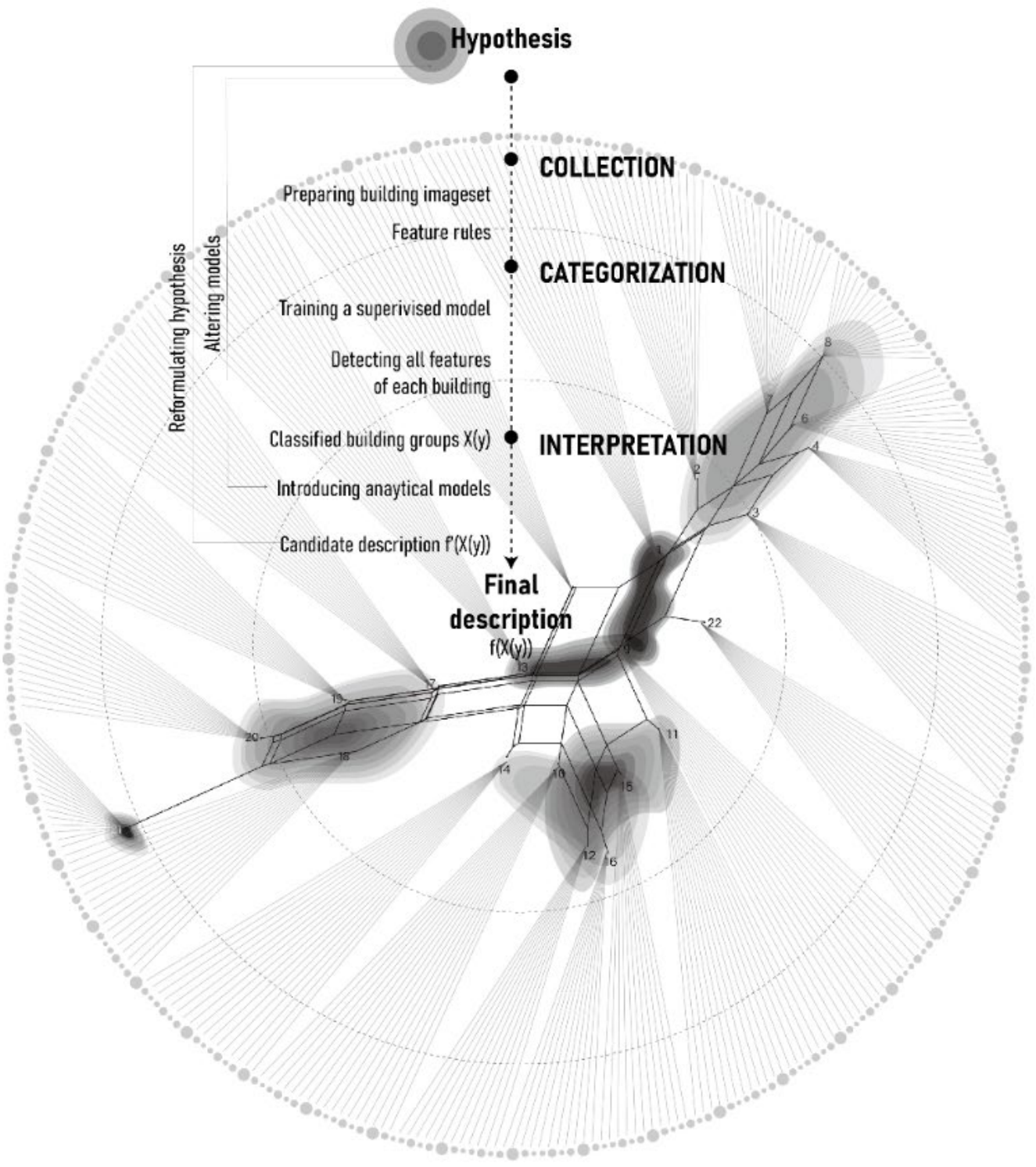

**Figure 1.** A visual intelligent framework of vernacular architecture

## 3. The Case of Singapore Shophouses

Built between the 1840s and 1960s, the Singapore shophouse is “a vernacular building of two to five-storey high; following a basic floor plan of a narrow frontage and common party-walls; with a versatility of being able to serve many functions such as economic, residential and recreational purposes”[27] (Figure 2). Shophouses constituted much of the urban fabric in pre-World War II Singapore[28] but rapid urban renewal since the 1960s has demolished many in the city. It was not until the late 1980s that the Urban Redevelopment Authority (URA) recognised shophouses as artefacts of a nation’s cultural identity and proposed to preserve shophouses. Currently, there are about 6,500 shophouses included in the conservation scheme, and those in the four historic districts of Boat Quay, Chinatown, Kampong Glam, and Little India are regulated with the strictest conservation rules that the entire building envelope is to be retained and restored[29].

The façade assumes a pivotal role in organising information about shophouses and is frequently singled out for structuring formal taxonomies of shophouses. Most prominently are the **six styles** proposed by the URA, namely early shophouse style (1840-1900), first transitional shophouse style (1900-1910), late shophouse style (1900-1940), second transitional shophouse style (late 1930s), art deco shophouse style (1930-1960), and modern shophouse style (1950-1960) (see Supplementary Information, ‘Six Styles of Singapore Shophouses by Urban Redevelopment Authority at Singapore’)[30].

Upon detailed scrutiny, we contend that this taxonomy raises concerns that are at odds with the perspectives of various other studies[31]. Firstly, to name some styles as ‘transitional’ is problematic since they never showcase an explicit intention of transiting to something new. Secondly, URA places the 'second transitional shophouse' after the 'late style shophouse'. This sequence differs from the sequence proposed by cultural geographer Victor Savage, who places the 'second transitional shophouse' before the 'late style shophouse' [32]. It also differs from Y.W. Tan’s observation on the sequential order of Penang shophouses[33]. Lastly, the history of shophouses dates only around

---

[27] Victor Savage, “Singapore Shophouse: Conserving a Landscape Tradition,” *SPAFA Journal* 11, no. 1 (2001): 6.
[28] National Heritage Board, “The Singapore Shophouses: Defining Early Singapore’s Cityscape,” Roots, June 21, 2002, https://www.roots.gov.sg/en/stories-landing/stories/singapore-shophouses/story.
[29] Urban Redevelopment Authority, “Conservation Principles,” Urban Redevelopment Authority, August 4, 2023, https://www.ura.gov.sg/Corporate/Guidelines/Conservation/Conservation-Guidelines/Part-1-Overview/Introduction.
[30] Urban Redevelopment Authority, “Understanding the Shophouse,” Urban Redevelopment Authority, August 4, 2023, https://www.ura.gov.sg/Corporate/Guidelines/Conservation/Conservation-Guidelines/Part-1-Overview/Understanding-The-Shophouse.
[31] Savage, “Singapore Shophouse: Conserving a Landscape Tradition”; Jon Sun Hock Lim, “Colonial Architecture and Architects of Georgetown (Penang) and Singapore, between 1786 And1942,” *CCK BATCHLOAD 20200605* (Ph.D Thesis, Singapore, National University of Singapore, 1991), https://scholarbank.nus.edu.sg/handle/10635/169028; Kip Lin Lee, “Identifying the Shophouse,” in *Singapore Chinatown Conservation* (Singapore: Singapore Tourism Board, 1985).
[32] Savage, “Singapore Shophouse: Conserving a Landscape Tradition.”
[33] Savage; Yeow Wooi Tan, *Penang Shophouses: A Handbook of Features and Materials* (Tan Yeow Wooi Culture & Heritage Research Studio, 2015).

120 years, and the emergence of different styles of shophouses has not always been seamlessly one after another, or diachronic. Jon Lim's research shows that architects like Wan Mohamed Kassim and Yeo Hock Siang had already submitted building drawings with spectacular elevations before 1900 for approval by government authorities[34]. These cases do not fit well with the current stylistic classification. A similar situation is observed in Emerald Hill, where according to architectural historian Lee Kip Lin "the age of the older houses which span between 1901 to 1918, the façade treatment var(ies) considerably in detail." [35] Lee in his article 'Identifying the shophouse'[36], opposes the categorisation of shophouses by time period. An inclusive formal taxonomy of Singapore shophouses may not necessarily adhere to a chronological principle. Identifying 'key styles' serves, therefore, as a heuristic device for understanding Singapore shophouses but can hardly cover their myriad variations[37].

This study uses a digital-driven research framework to guide the formal taxonomy in order to provide a deeper understanding of shophouses. We hypothesise that **there is an undiscovered relationship between the facade form and the cultural meaning of shophouses (h1).** As suggested in the framework, this hypothesis is preliminary and will be gradually refined and further developed in subsequent steps. We selected Chinatown as our study site, the largest historic district in Singapore, with 1,277 shophouses gazetted for conservation. Data are defined as images that describe the shophouse frontage.

## 3.1. Collection

We collected ***building frontage images*** from two channels, one from Google Street View (GSV) and the other from on-site photography. This strategy is not pre-designed but results from iterative experiments: a 100% reliance on the public geospatial database proves impractical in our case. This is because the maximum resolution of GSV images that can be output currently is 640*640, which is relatively low for completing the entire data analysis procedures. As such, we used the GSV image set for the feature verification task for building façades. Concurrently, we also introduced on-site photography, with an increased image resolution of 3024*4032, to deal with feature detection with respect to specific façade elements.

All images are captured with the camera positioned to face each building's centroid, with the line between the camera and the centroid set perpendicular to the shophouse frontage. The prepared two frontage image sets comprising GSV and on-site photos cover all shophouses in Chinatown as well as those in three other historic districts, including Boat Quay, Kampong Glam and Little India, totalling **3,103** images. The dataset covering a much larger area than the study site benefits the model training in the following stage. It enables the model to perform more accurately in categorisation tasks, effectively reducing human labour.

[34] Lim, "Colonial Architecture and Architects of Georgetown (Penang) and Singapore, between 1786 And1942."
[35] Lee, "Identifying the Shophouse."
[36] Lee.
[37] Upton, "The Power of Things," 273.

Concurrently, we set the **feature rules.** The 'feature' is the key to describing the change in building forms. Referring to our previous hypothesis, it can also be regarded as the unit that transmits cultural meanings[38]. Because it is formulated manually by researchers, one has to note that the scale of the feature may greatly influence the results of formal taxonomy. Ensuring that the feature unit has a relatively stable form is thus important. In our case, the feature rules are set on the scale of façade elements. This is because each façade element on the shophouse has a relatively constant formal character and an explicit similarity with European classical, Chinese traditional, Malay traditional, and modern architectural vocabularies. Each shophouse façade can be read as the composition of varied façade elements. Extended data 1 provides a glossary of the Singapore shophouse façade we formulated through extensive literature review and field study.

Façade elements are either decorative, functional, or semi-decorative-semi-functional. The frequency of these façade elements emerging on the shophouse frontage also varies; some are frequently found, like festoon, *bian-e* (Chinese caligraphy), and *kepala tingkap* (Malay transom); some others are less seen, like the Diocletian window, *ge-shan* (Chinese partition screen), and *tangga batu* (stone staircase). This glossary demonstrates that the facade language of Singapore shophouses is influenced by different ethno-cultural traditions. It also serves as preliminary evidence to support the initial hypothesis (h1). However, while surfacing at this stage through human interpretation of the glossary, this argument is not encoded into the following categorisation phase automated by machine.

## 3.2. Categorisation

The categorisation phase involved ***tallying the façade elements on each building*** within the Chinatown study site based on the glossary. To achieve greater efficiency and objectivity, we trained a supervised model to auto-detect the architectural elements on the shophouse frontage. The working flow is detailed in Figure 2 beginning with the setting up of label classes of façade elements. These label classes were filtered out from the glossary of the Singapore shophouse façade based on two principles. First, façade elements with insufficient instances (n<40), indicating relative unimportance[39], were removed. Second, we trained a site-specificity model by linking 3,103 GSV images with their situated historic districts. Using a deep learning model explanation method, Grad-CAM (Gradient-weighted Class Activation Mapping), we concluded that 'featured architecture elements above the ground floor' are the key to identifying shophouse facades. In the end, we defined 14 label classes of façade elements: main pilaster, fanlight, secondary pilaster, festoon, modillion, Chinese plaque, Chinese decorative panel, Malay transom, fretwork fascia, majolica tiles, long window, modern window, shades, stepping parapet. We also introduced the building label to focus the learning area to within the upper-level of the shophouse façade.

[38] Richard Pocklington, "What Is a Culturally Transmitted Unit, and How Do We Find One?," in *Mapping Our Ancestors*, ed. Carl P. Lipo et al. (New York: Routledge, 2006), 19–22; Michael J. O'Brien and R. Lee Lyman, "Darwinism and Historical Archaeology," in *International Handbook of Historical Archaeology*, ed. David Gaimster and Teresita Majewski (New York, NY: Springer, 2009), 227–52, https://doi.org/10.1007/978-0-387-72071-5_13.
[39] "Darwinism and Historical Archaeology," 233.

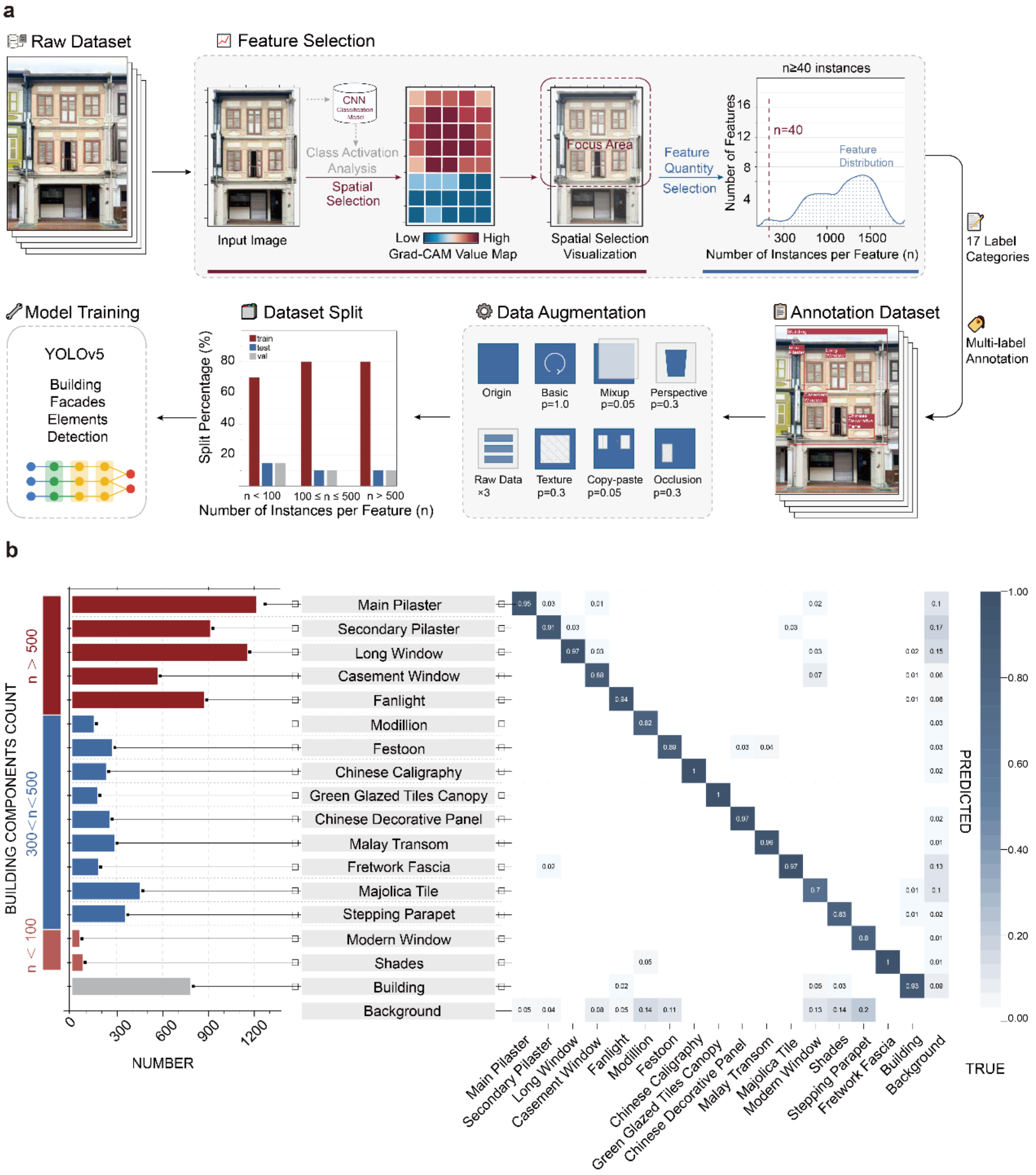

**Figure 2.** The collection and categorisation of Singapore shophouse façades. a, the workflow from collection to categorisation. b, the performance of the YOLOv5-based detection model in identifying the 15 label classes

Next, we labelled 708 out of 3,361 images from on-site photography with 15 label classes (14 label classes plus the building label) and divided them into training, validation, and testing. Since multiple elements are often labelled in each image, the number of labelled elements far exceeds the number of images. **Figure 2b** shows the number of samples by class, presenting an uneven distribution. We implemented a frequency-stratified splitting strategy to ensure robust model performance across all architectural elements while respecting their natural distribution patterns. For elements with fewer than 200 instances, which often represent distinctive architectural features, we adopted a 70:15:15 (train: validation: test) ratio to ensure sufficient evaluation samples. For more frequent elements, we maintained the conventional 80:10:10 split ratio. We employed data augmentation techniques and class weight adjustments to enhance the model's further learning on less frequent elements. This comprehensive approach enabled our YOLOv5-based detection model to achieve an overall accuracy of 0.94332 on the 10% test set, demonstrating robust performance across both common and rare architectural elements.

We employed the model to identify the presence or absence of 14 traits in 1,277 shophouses in Singapore's Chinatown. The results produced a 14-digit binary string for each building, where 0 indicates absence and 1 indicates presence. Out of 1,277 buildings in Chinatown, 60 were identified by the model as having no feature elements and were represented by a 14-digit string of zeros. Undergoing a human-aided inspection, we confirmed 56 shophouses as variants. Extended data 2 shows a list of shophouse variants. For the rest of the 1,211 shophouses, by counting duplicates of their 17-digit strings and filtering out those with duplicate values less than 8, we concluded with 25 string labels, each representing one type of shophouse façade with a specific element composition (Figure 3a).

**Figure 3c** shows the correlation index of façade elements in composing the main types. Majolica tiles and Malay transom are excluded from the 25 main types of compositions, indicating their very low presence in Chinatown. The other 12 elements, however, automatically aggregated into several clusters. For example, three elements – stepping parapet, modern windows and shades – correlated frequently. However, their co-emergence with other elements is low. **Figure 3b** shows the frequency and percentage of shophouses containing each façade element. Long windows and main pilasters are the most prevalent façade components, found in 81% and 82% of the buildings, respectively. In historical archives, the two elements were also noted as the earliest found in shophouses, a prominent example being Baba Yeo Kim Swee’s godowns, constructed in the 1840s. They were designed by George Coleman, Singapore’s first professionally trained architect with classical architecture education. Although these buildings are devoid of ornamentation, the modular repetition of their two principal elements along the row produces a sense of classical order, which appealed to the colonial authorities of the day.

Fanlights and secondary pilasters are found in 52% and 28% of the buildings respectively. Chinese calligraphy and Chinese decorative panels are found in 15% and 12% of the buildings respectively. Majolica tiles, Malay transom, and fretwork fascia registered the lowest percentages, 0.1%, 1.6%, and 0.18% respectively. The results indicate that the influence of the Malay is weak in an ethnic district predominated by the Chinese. However, being a British colony, the influence of European classical architecture dominates the urban vernacular landscape. The analysis of building elements can therefore help suggest the factors driving building and architectural changes but this perspective remains very preliminary.

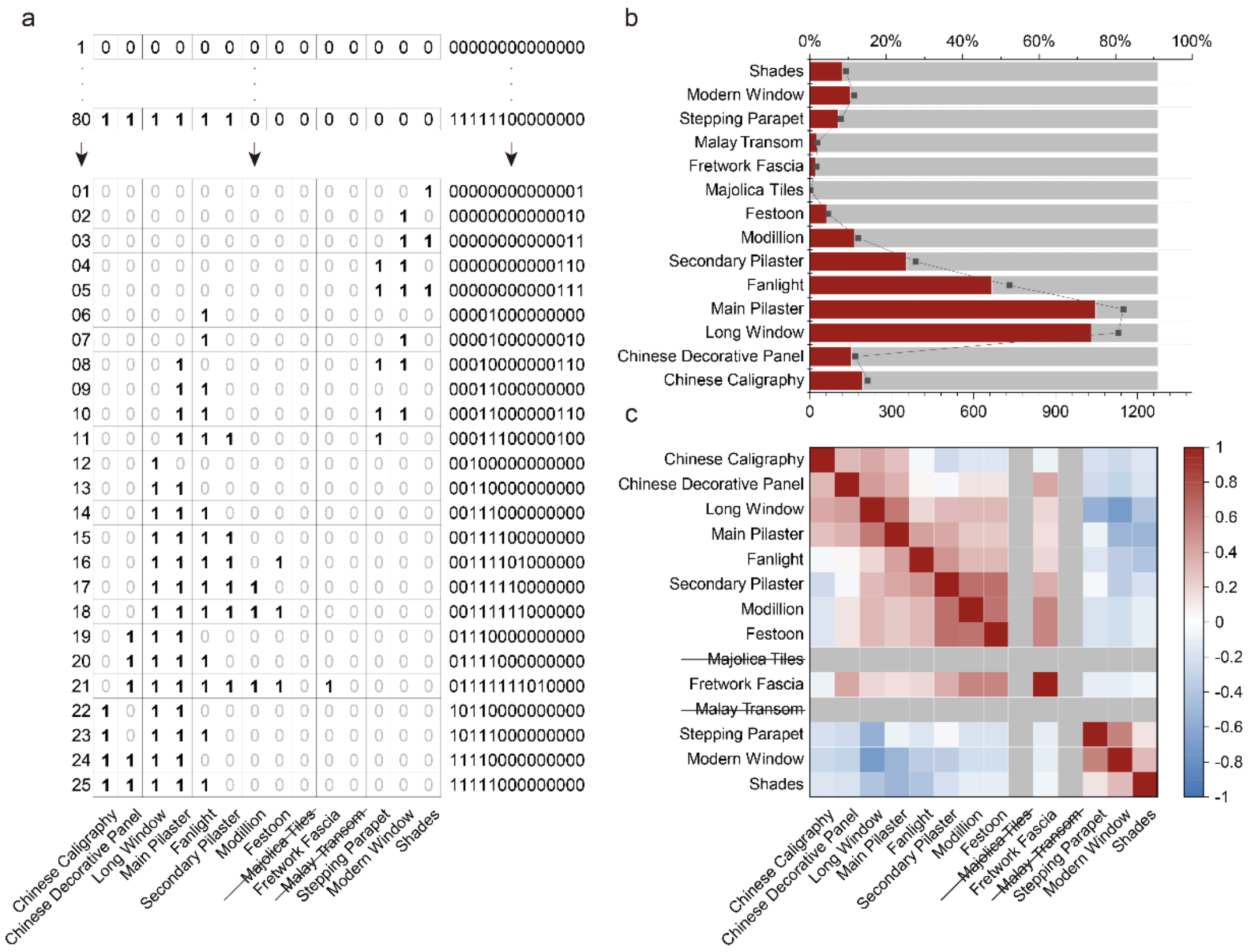

**Figure 3.** The categorisation results of Singapore shophouse façades. a, the 25 types of Singapore shophouse facades in Chinatown. b, the correlation index of façade elements in composing main types. c, the frequency and percentage of shophouses containing each façade element.

## 3.3. Interpretation

To unpack whether and how one shophouse façade type is related to another, we must introduce statistical models with cultural hypotheses to model the change in building forms. Notably, the field of vernacular architecture has never developed a statistical model of its own. The explanations of vernacular architecture based on qualitative observation are varied but sometimes contradictory[40]. Validating these explanations through statistical models is necessary but also challenging, as it requires encoding cultural hypotheses specific to vernacular architecture into the model. Currently, we can only depend on statistical models applied to sociology and archaeology. Despite not being intended for vernacular architecture studies, some of these models share a hypothesis similar to that of the field of vernacular architecture, like cultural evolution and cultural diffusion[41].

[40] Lawrence, "The Interpretation of Vernacular Architecture."

This study employs a phylogenetic method to relate 25 Singapore shophouse types. Cultural phylogenies are built on the proposition that ***formal change of artefacts results from cultural transmission***. This proposition is a further development of our preliminary hypothesis (h1) and is proposed as research findings increasingly emerge. We retrieved three hypothetical graphic patterns from existing phylogenetic models to try to explain the formal change of vernacular architecture: ***line*** (h1-1), ***tree*** (h1-2) and ***network*** (h1-3). The three patterns are split from the cultural phylogenies, resulting in competing views (Figure 4), and we elaborate on each in the subsequent discussion.

### 3.3.1. *Line*

The linear model, commonly known as seriation, suggests "arrang(ing) comparable units in a single dimension (that is, along a line) such that the position of each unit reflects its similarity to other units."[42] Seriation assumes cultural transmission is one-way. This one-way transmission can be further incorporated with temporal dimension to describe a cultural evolutionary mode characterised by heritable continuity. An essential characteristic of this heritable evolution is that "the parental taxon goes extinct when the daughter taxon appears"[43].

The technique has been used on some informal occasions to date architecture based on the presence and absence of certain architectural features[44]. Notably, the heritable order can only be established if the façade elements are secured as reliable chronological indicators. Given the relatively short history of Singapore shophouses, we speculate that heritable continuity can be manifested in part of the 14 elements but not all of them.

**Figure 4** shows that 25 types have been seriated based on their composed façade elements in a non-constrained environment. Results specify a signal of heritable evolution emerging between the former nine elements of Chinese calligraphy – Chinese decorative panel, long window, main pilaster, fanlight, secondary pilaster, modillion, festoon, fretwork fascia, and the latter three elements of stepping parapet, modern windows and shades – resulting in three distinct groups: 'Group traditional' (24, 22, 19, 23, 13, 20, 14, 09, 15, 17, 06, 16, 21, 18), 'Group Transitional' (11, 10, 07, 08), and 'Group Modern' (4, 2, 5, 3, 1). Historical archives support that evolutionary trend. Traditional shophouses built before and during the early 20th-century feature ornaments with ethnocultural traditions. However, the development of new materials and technologies in the 20th century have led to a rethinking of tradition and a preference for streamlined ornamentation. Entering the 1930s, modernism took hold amidst an economic recession and the influence of global architectural trends. Modernist architecture rebelled against tradition, viewing ornamentation as a crime. The principle of “form follows function” dominated, incorporating geometric forms with functional elements, presenting a transformative look. Seriation analysis reveals a chronological narrative hidden in the 25 main types by concluding on the three groups as noted above.

[41] O'Brien and Lyman, "Darwinism and Historical Archaeology."
[42] William H. Marquardt, "Advances in Archaeological Seriation," *Advances in Archaeological Method and Theory* 1 (1978): 258.
[43] R. Lee Lyman and Michael J. O'Brien, "Seriation and Cladistics: The Difference between Anagenetic and Cladogenetic Evolution," in *Mapping Our Ancestors: Phylogenetic Approaches in Anthropology and Prehistory*, ed. Carl P. Lipo et al. (New York: Routledge, 2006), 66.
[44] Marquardt, "Advances in Archaeological Seriation," 259.

However, a closer look at the neighbour relationship between two off the seriated types reveals a ‘problem’. For example, Type 24, composed of four elements of Chinese calligraphy, Chinese decorative panels, main Pilasters, and long windows, is designated as the earliest type but this does not align with historical archives. This may be explained by the fact that the linear mode has greatly simplified the cultural narrative of Singapore shophouses.

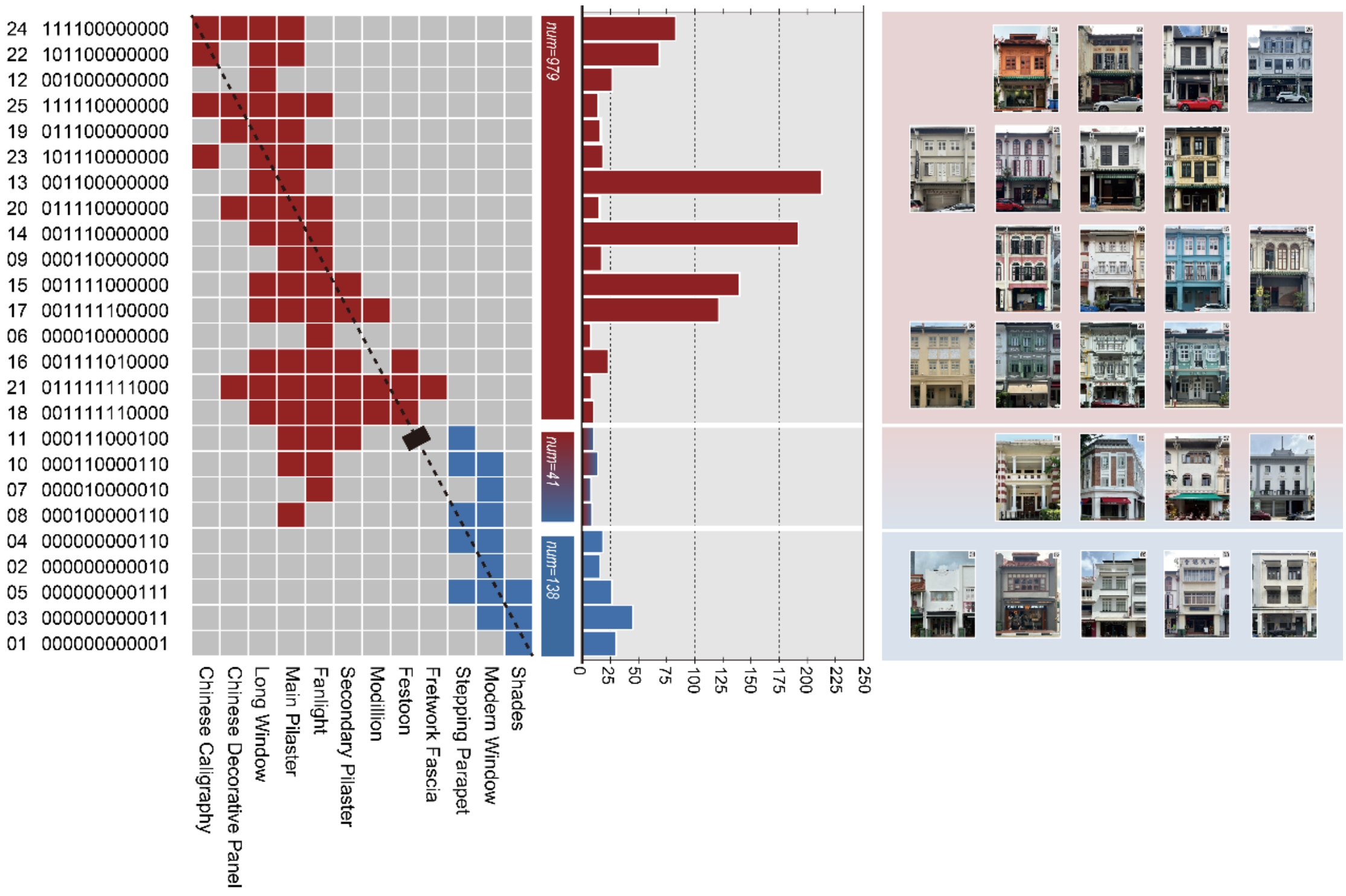

**Figure 4.** The interpretation of Singapore shophouse facades based on the line mode.

### 3.3.2. *Tree*

The tree mode, developed upon the neighbour-joining algorithm, suggests a phylogenetic tree to depict the evolutionary relationship among the taxas. This tree assumes a branching pattern of lineage splitting to model cultural transmission. When a lineage splits, it divides into two distinct groups. Like the linear mode, the tree mode also adheres to the principle of heritable continuity from ancestors to descendants.

Despite not using the computational method, architectural historians have long employed the tree mode to trace the genealogy of architecture. Examples include Sir Banister Fletcher's Tree of Architecture[45] and Liang Ssu-Ch'eng's drawing on the Evolution of Types of the Buddhist Pagoda[46]. These studies exemplify the applicability of the branching model in modelling the evolutionary history of architecture. Notably, the tree mode of cultural transmission is built upon the hypothesis that "similarities and differences among cultures are the result of a combination of predominantly within-group information transmission and population fissioning"[47]. Given the context that modern Singapore was born out of the colonial trade network of Imperial Britain in the 19th century, we speculate that the evolution of the Singapore shophouse must be influenced by the constant flow of people and social materials into the island.

**Figure 5** illustrates the phylogenetic tree of the 25 shophouse facade types. According to the hypothesis of cultural evolution, each internal node of the tree represents a heritable evolution event that "the parental taxon goes extinct when the daughter taxon appears"[48]. When viewed as a whole, the tree has three prominent clades, containing types with predominantly modern elements, European elements, and Chinese elements respectively. The phylogenetic tree suggests the Clade Blue and Orange split from the same node, suggesting they evolved from a common ancestor ($A_{bo}$), while the Clade Red and ancestor $A_{bo}$ separated from a different node, indicating they evolved from another lineage. Each clade is composed of many subclades. The smallest clade always includes two types split from the same node. They are considered sister groups, indicating the highest cultural proximity (3 and 5, 8 and 10, 1 and 6, 18 and 21, 24 and 25, 22 and 23, 19 and 20).

Based on our direct observations and historical archives, the smaller clades in the tree offer greater explanatory power for types in which most of the compositional elements are repetitive. For example, types 3 and 5 reveal very similar visual language as well as building crafts, indicating both types are formulated by architects based on a particular prototype with minor revision. However, larger clades with multi-layers of evolution in the tree are less explanatory. This is because phylogenetic trees follow a strict bifurcating model. In referring to historical archives, many of the ancestor-descendant relationships indicated by the phylogenetic tree are invalid. Most prominently, no evidence shows that the red clade featuring predominantly Chinese elements emerged earlier than the orange clade featuring predominantly European elements. We conclude that the tree mode cannot fully address the evolution of

[45] Banister Fletcher and Banister F. Fletcher, *A History of Architecture on the Comparative Method for the Student, Craftsman, and Amateur* (New York: Charles Scribner's Sons, 1905), http://archive.org/details/historyofarchite00flet.
[46] Ssu-Ch'eng Liang, *Pictorial History of Chinese Architecture: A Study of the Development of Its Structural System and the Evolution of Its Types* (Cambridge, Mass.: MIT Press, 1984).
[47] Mark Collard, Stephen J. Shennan, and Jamshid J. Tehrani, "Branching versus Blending in Macroscale Cultural Evolution: A Comparative Study," in *Mapping Our Ancestors: Phylogenetic Approaches in Anthropology and Prehistory*, ed. Carl P. Lipo et al. (New York: Routledge, 2006), 53.
[48] Lyman and O'Brien, "Mapping Our Ancestors," 66.

Singapore shophouses.

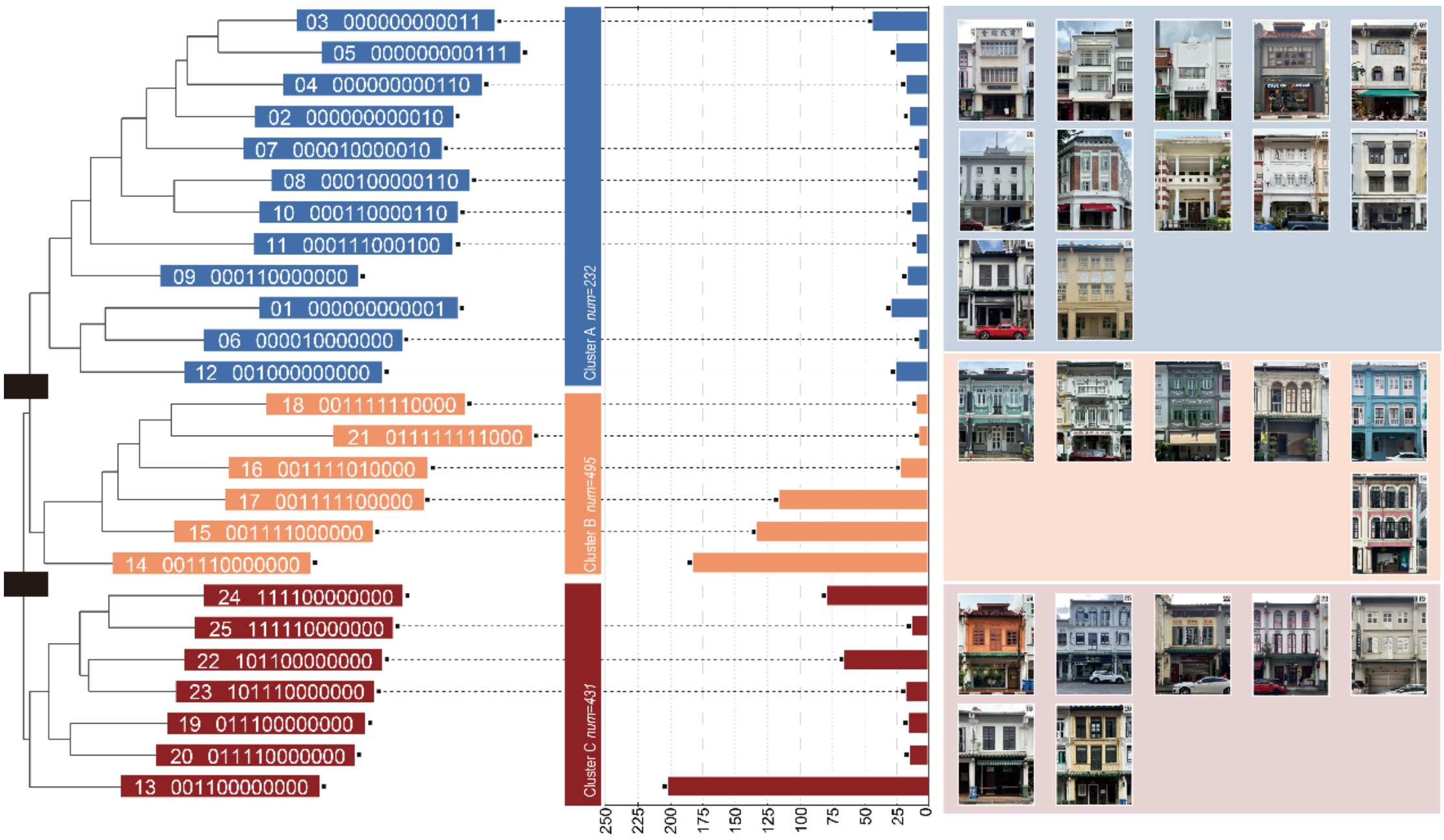

**Figure 5.** The interpretation of Singapore shophouse facades based on the tree mode.

### 3.3.3. *Network*

The network model, developed upon the neighbour-net algorithm, suggests a split graph to depict the relatedness among the cultural taxes. Unlike the tree model, which assumes binary evolution through hierarchical ancestor-to-descendant inheritance, the network model suggests a reticulate transmission, allowing multiple connections. These better capture non-hierarchical signals in cultural transmission, like cultural hybridisation, borrowing and parallel evolution. In this sense, the network model can best represent the hypothesis of cultural diffusion that "the patterns of similarity and difference among cultural assemblages are a consequence primarily of individuals in different groups copying each other's practices, exchanging ideas and objects, and marrying one another. "[49] Building on preliminary studies of façade elements that suggest frequent intra-ethnic exchange in Singapore's society, we speculate that a network model more effectively captures the relationships among the 25 types.

Notably, while the cultural diffusion hypothesis in social science was proposed in opposition to cultural evolution, the mathematically formulated network model does not discard the tree model but builds on it, enhancing its applicability. When interpreting the graph, a network resembles a tree, indicating a clear hierarchical evolutionary pattern; if the network contains loops, it suggests conflicting signals, such as hybridisation or multiple influences.

---

[49] Collard, Shennan, and Tehrani, "Branching versus Blending in Macroscale Cultural Evolution," 54.

**Figure 6** illustrates the phylogenetic network of 25 facade types. When viewed as a whole, the network exhibits a branching growth pattern extending outward from the centre toward distinct directions; the types in the centre have fewer featured façade elements. In contrast, those in the peripheral area exhibit more facade elements indicating the possibility of cultural evolution. We identified four clusters in the network: the Group Plain in grey, the Group Chinese in red, the Group Euro in yellow and the Group Modern in blue. The Group Plain comprises Types 12 and 13, featuring basic elements of long windows and main pilasters. Historical archives also suggest these two types were the first to emerge in Singapore, dominating the city's urban landscape in the early 19$^{th}$ century. Viewed in conjunction with the branching growth pattern, Group Plain can also be seen as the earliest-evolved entity. Group Chinese in red comprises six types with Chinese elements, featuring elements like Chinese calligraphy and Chinese decorative panels. Historical archives suggest that these types in Group Chinese emerged later than Group Plain and were initially constructed for affluent Chinese owners.

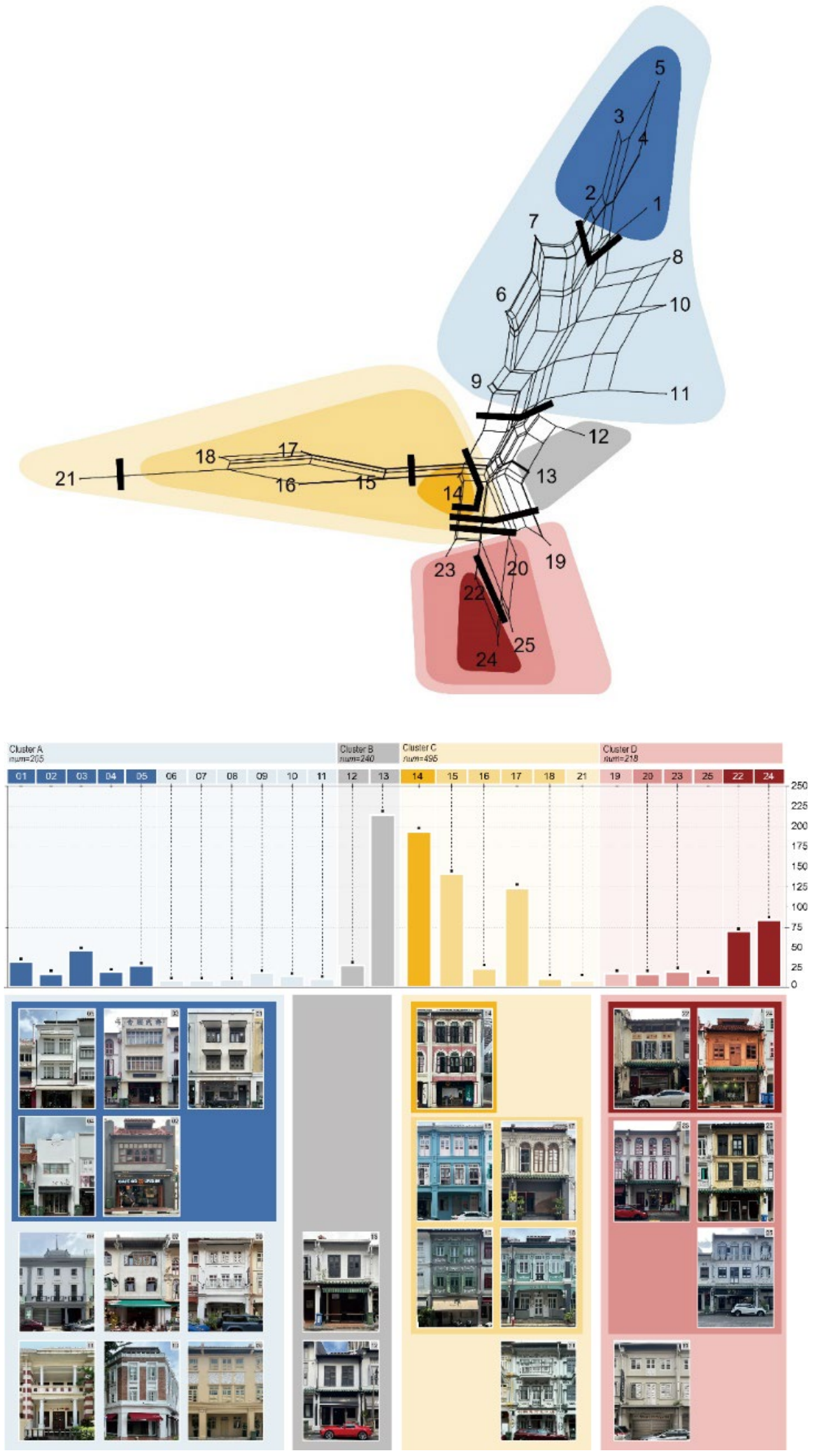

**Figure 6.** The interpretation of Singapore shophouse facades based on the network

mode.

Group European comprises six types with elaborate European elements, like fanlight, secondary pilaster, modillion, and festoon. Historical archives suggest these types also emerged later than Group Plain but proximately in the same period as Group Chinese. Since the later 19$^{th}$ century, the construction amount of Group Yellow surged in Chinatown, even exceeding Group Red. They were constructed under the commission of not only European owners but other ethnicities, reflecting a wider influence of colonial power in the society. Group Modern in blue comprised 11 types with modern elements, like stepping parapets, modern windows and shades. Historical archives suggest Group Modern emerging as the latest in the 20$^{th}$ century. Despite comprising 11 types, the total number of constructions in Group Blue in Chinatown is low. The economic downturn and disruptions of World War II significantly slowed new construction. Moreover, there is a dearth of heritable continuity between Group Modern and earlier Group Chinese and European. Viewed in conjunction with the branching growth pattern, Group Modern, like Group Chinese and European, can all be seen as having evolved from Group Plain. Comparatively, the evolutionary directions of Group Chinese and European are closer in terms of distance, while that of Group Modern is distinctively different.

Besides its tree-like structure, the phylogenetic network contains many loops, indicating conflicting transmission signals. Each group and type receive influence from multiple traditions, which complicates the categorisation informed by the tree. Nine distinct subgroups were further identified based on the four primary groups, which are also the nine styles that best capture the character of Singapore shophouses: **Plain Style** (13, 12), **Plain Chinese Style** (19), **Chinese Style** (22, 24), **Chinese-European Style** (20, 25, 23), **Plain European style** (14), **Elaborated-European Style** (15,17,16,18), **Ethnic-hybrid Style** (21), **Stripped Ethnic Style** (9, 11, 6, 10, 7, 8), and **Modern Style** (1, 2, 3, 4, 5) **(Figure 7).** Plain Style overlaps with Group Plain in grey. It is primarily function-oriented, with minimal ornamentation: a modest amount of moulding serves as capitals for the columns and pilasters, while the windows are square-headed and fitted with jalousie shutters.

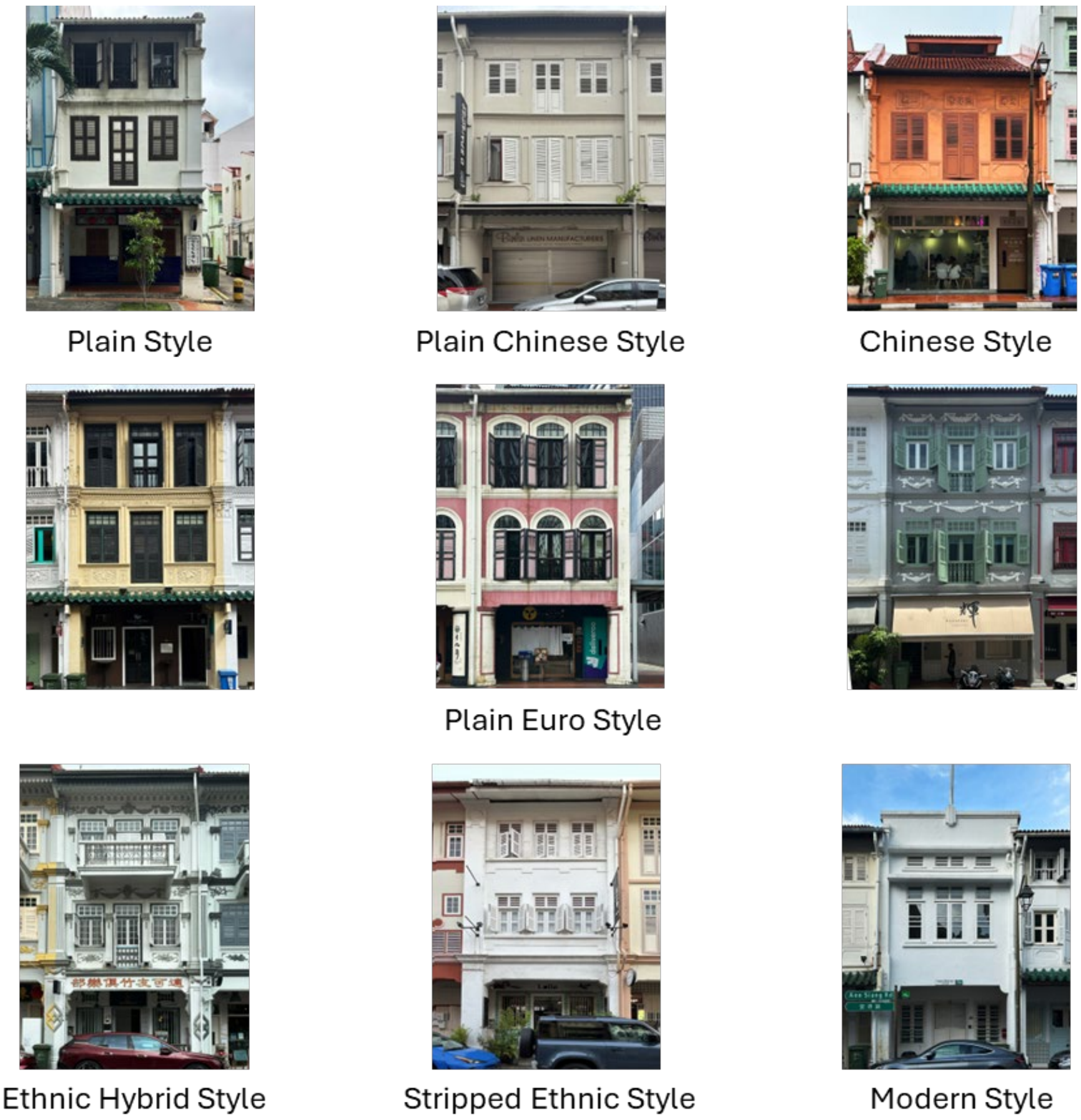

**Figure 7.** Nine shophouse styles in Singapore's Chinatown

Plain Chinese Style, Chinese Style and Chinese-European Style are all split from Group Chinese in red. Chinese Style is defined by traditional Chinese decorative elements, either wall plaques with auspicious characters, plaster relief panels with Chinese motifs, or a combination of both. Chinese-European Style, however, incorporates these traditional Chinese decorative elements with classical European elements, creating a fusion look. Historical studies indicate these Chinese-European Style shophouses were primarily built by Chinese merchants who wanted to preserve their cultural roots while aligning with colonial elites to enhance their social and economic status[50]. The style exemplifies the cultural diffusion from Group European to Group Chinese.

The Plain European Style, Elaborated European Style and Ethnic-hybrid Style are all split from Group European in yellow. The Plain European Style is similar to Plain Style but with added European fanlights. The Elaborated European Style, developed from the Plain European Style, features more decorative European elements such as secondary pilasters, modillions, and festoons. The Ethnic-Hybrid Style combines traditional Chinese, European and Malay elements, presenting a peculiarly vernacular mix. This explains Type 21, which has an exceptionally long line extending from the rest of the European Group. The Ethnic-Hybrid Style shares many similarities with the Chinese European Style within Group Chinese, as both represent forms of cultural hybridity. However, the Ethnic-Hybrid Style also displays a formal language dominated by a decidedly European classical language.

[50] Ronald Knapp, *Chinese Houses of Southeast Asia: Eclectic Architecture of the Overseas Chinese Diaspora* (Tokyo, 2010).

Historical archives reveal a more nuanced understanding of the cultural transmission of ethnic traditions as reflected by the Ethnic-Hybrid Style. Statistically, there are eight Ethnic-Hybrid Style shophouses in Chinatown, all located at Bukit Pasoh Road. Historical architectural drawings suggest that they were all designed by Swan & Maclaren, one of the most renowned architectural firms in Singapore since the colonial period, employing professional architects trained in European classical architecture. The owner of this row of eight shophouse is Seow Chwee, a Strait-Chinese merchant. In fact, the initial design by Swan & Maclaren showcased an orthodox classical architectural image. Alternating triangular and segment pediments were employed to crown the window, while the decorative panels beneath them featured festoon ornamentation **(Figure 8)**. However, their design was not fully reflected in the final façade and could have been modified by the Chinese artisan under the owner's commission, which was a common practice during the time[51]. The festoon on the second floor was replaced by eight Chinese symbols with auspicious meanings, the Covert Eight Immortals. They also added the fretwork fascia of the traditional Malay house under the roof edge. The Ethnic-Hybrid Style is the most elaborate among the 25 types. It is also the only case that incorporates Malay elements in Chinese-owned buildings. Despite Type 21 exemplifying multi-ethnic integration, historical archives reveal that ethnic competition also played a significant role in shaping its hybrid style.

[51] Julian Davison, *Swan and Maclaren: A Story of Singapore Architecture* (ORO Editions & The National Archives of Singapore, 2020), 269.

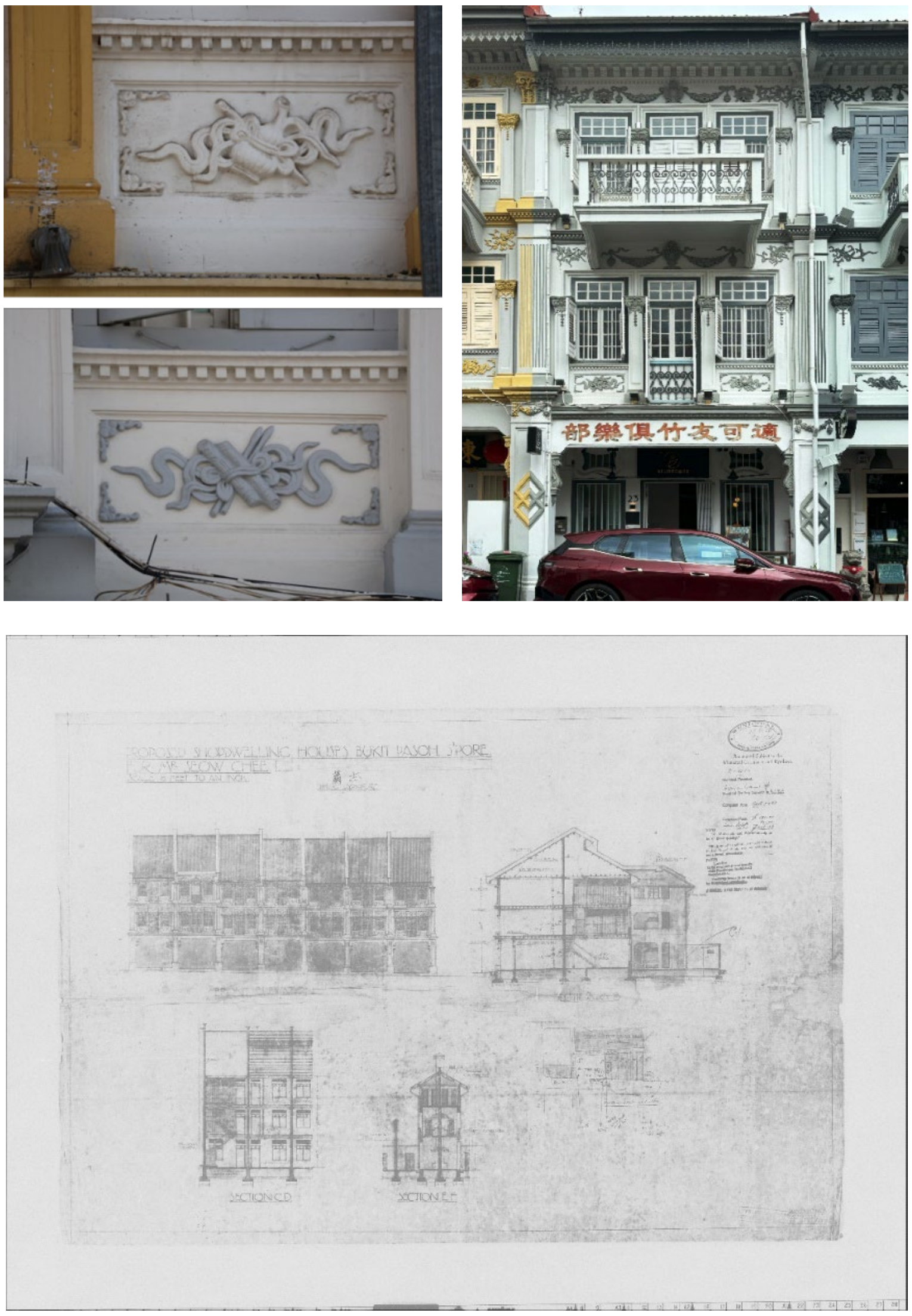

**Figure 8.** The architectural drawing submitted by Swan & Maclaren in 1926 for a row of shophouses at Bukit Pasoh Road.

The Stripped Ethnic Style and Modern Style are both split from Group Modern in blue. While the Stripped Ethnic Style reduces traditional ornamentation but incorporates modern geometrical elements, leading to some innovative forms, the Modern Style features only modern geometrical elements like stepping parapets, modern windows, and shades. The Stripped Ethnic Style occupies a large area of the phylogenetic network, indicating that the six types are relatively distant. This is because, driven by the manifesto of anti-tradition, architects in the early 20$^{th}$ century were keen on exploring a new visual language that could better embody the spirit of the industrial era. The Stripped Ethnic Style was born out of this context, showcasing great diversity and a spirit of experimentation. As the Modern Style emerged, the experimental exploration stabilised.

The phylogenetic network provides a critical perspective for examining Singapore shophouses featuring multi-ethnic characters. The exchange of ethnocultural traditions occurred during a period of economic prosperity from the late 19th century onwards. However, ethnic power dynamics remained unbalanced, as reflected in the dominance of European classicism in architectural styles. Statistically, European influences prevailed, with the Elaborated-European and Plain European Styles accounting for 23.57% and 17.23% of the total, respectively. Chinese influence was comparatively lower, with the Chinese Style comprising 13.63%. Fusion styles were the least common, with the Chinese-European, Ethnic Hybrid, and Stripped Ethnic Styles representing 4.54%, 3.52%, and 4.70% of the total, respectively. This distribution suggests that different ethnic groups remained in competition, displaying a pattern of parallel evolution rather than direct convergence. While some degree of architectural fusion did occur, it remained limited, reinforcing the prevailing cultural distinctions in Singapore.

## 4. Conclusion

In this study, we have proposed a research framework of utilising vision-intelligent technologies to unveil the genealogy of vernacular architecture. We suggest viewing the interpretation of historical artefacts as a process of finding a function $f$ from a discrete data set. With this, visual intelligence technologies can be assembled to support researchers in four key aspects: efficiently mining building data; converting data into feature clusters; identifying the mathematic function that best relate these clusters; and converting that function into an architectural narrative. By recasting an architectural issue as a mathematical question, a computational framework to vernacular architecture can break free from the reductive narrative models of traditional studies. It probes more deeply into the synchronic threads of development embedded within diachronic processes, promising a breadth and depth that conventional methods have yet to achieve.

Grounded in an epistemological stance of solving function, we formulate a systematic workflow that incorporates data collection, feature clustering, model construction and interpretation with spatial data acquisition, computer vision, and scientific data analysis technologies. First, spatial data acquisition technologies have significantly improved the quality of primary data for vernacular architecture studies. Different from traditional methods that rely on piecing together fragmented materials to understand architecture, spatial data acquisition techniques enable large-scale, multi-temporal building representation with systematic data. Second, the incorporation of machine learning – especially computer vision and deep learning – has ushered vernacular architecture research into a new stage of automation in feature extraction and feature cluster construction.

Traditional studies have conventionally depended on experts manually identifying features based on their experience. In contrast, machine learning algorithms can automatically ‘pick up’ morphological characteristics from vast amounts of architectural images and structural data. This shift from individual expertise to a data-driven approach, as explained in the introduction can help to uncover deeper, more nuanced patterns.

Finally, developing mathematical functions that capture the relatedness among feature clusters is key to unveiling the genealogy of vernacular architecture. A common pitfall in many vernacular architecture studies is overlooking the need for setting cultural hypotheses at the outset. As a result, research often becomes a linear narrative of progress or a fragmented functionalist framework, lacking depth. A vision-intelligent approach to vernacular architecture demands that scholars begin with clear hypotheses, aiming for both comprehensive, nuanced observations and straightforward, impactful concepts. The hypotheses also need to be constantly revisited and updated in view of empirical findings as they are uncovered, as our study shows.

We have demonstrated the effectiveness of the research framework through a case study of Singapore shophouses. Findings suggest that the formal change of shophouses across time and space is best captured by the phylogenetic network, which organises the types into nine distinct clusters and reveals concurrent signals of cultural evolution and diffusion. Here, cultural evolution manifests as a transition in styles from the traditional to the modern, while cultural diffusion – particularly during periods of economic prosperity – is characterised by the emergence of diverse ethnic styles. However, we also observe that different ethnic groups remained in competition in colonial Singapore, displaying a pattern of parallel evolution rather than direct convergence. While some degree of architectural fusion did occur, it remained limited, reinforcing the prevailing cultural distinctions.

Extending this framework beyond Singapore is principally a matter of calibrating three interchangeable modules—data acquisition, feature extraction, and phylogenetic inference—to the vernacular system at hand. Remote sensing tools like drone photogrammetry and satellite stereo-imagery can supply the large-area, multi-temporal coverage needed to document vernacular architecture. Domain experts would concurrently translate local building knowledge into an artefactual grammar. Once encoded, the same vision-intelligent pipeline can automatically extract these features, cluster them hierarchically, and generate a phylogenetic network that makes both diachronic (evolutionary) and synchronic (diffusionary) relationships legible. Deploying the model across disparate settings would not only test its robustness, but could also reveal cross-regional "deep homologies" in vernacular problem-solving, enriching comparative studies of cultural resilience and architectural adaptation worldwide.

Rather than directly using digital tools to generate cultural narratives, we focus on how these tools can be better assembled to guide scholars' intuition to yield fruitful and insightful findings. Our case study highlights the advantages of this paradigm in uncovering cultural phenomena that are too vast for researchers to discern patterns through traditional observation alone. We hope this framework serves as a valuable mechanism for incorporating digital tools into vernacular architecture studies, fostering more fruitful collaborations between architectural and computer science scholarship in the future.

## Acknowledgements

We thank P. P. Ho at the Department of Architecture, National University of Singapore, for his invaluable insights and constructive feedback on this research. We acknowledge funding from the Heritage Research Grant (HRG-057 to all authors) provided by the National Heritage Board in Singapore.

## Author contributions

X.X. conceived the project ideas and designed the research framework. X.X. and Z.T. collected the image and textual data. X.X., Y.Y. and Z.T. implemented the computational analysis in Python and SplitTree. X.X. interpreted the results and wrote the manuscript. Z.T., T.-C.C. and C.-K.H. were involved in the discussion and manuscript revisions.

## Competing interests

The authors declare no competing interests.